\documentclass[aps,prl,final,twocolumn,letterpaper]{revtex4}
\usepackage{amsmath}
\usepackage{graphicx}
\usepackage{hyperref}
\usepackage{braket}
\usepackage{xcolor}

\begin{document}

\title{On-Chip Laser-Driven Free-Electron Spin Polarizer}

\author{Clarisse Woodahl\thanks{Corresponding author: cwoodahl@stanford.edu}}
\email{cwoodahl@stanford.com}
\author{Melanie Murillo}
\author{Charles Roques-Carmes}
\author{Aviv Karnieli}
\author{David A. B. Miller}
\author{Olav Solgaard}
\affiliation{E. L. Ginzton Laboratory, Stanford University, Stanford, CA, 94305, U.S.A.}

\date{\today}

\begin{abstract}
Spin-polarized electron beam sources enable studies of spin-dependent electric and magnetic effects at the nanoscale. We propose a method of creating spin-polarized electrons on an integrated photonics chip by laser driven nanophotonic fields. A two-stage interaction separated by a free space drift length is proposed, where the first stage and drift length introduces spin-dependent characteristics into the probability distribution of the electron wavefunction. The second stage uses an adjusted optical near-field to rotate the spin states utilizing the spin-dependent wavepacket distribution to produce electrons with high ensemble average spin expectation values. This platform provides an integrated and compact method to generate spin-polarized electrons, implementable with millimeter scale chips and table-top lasers. 

\end{abstract}

\maketitle

\section{Introduction} 
The Stern-Gerlach (SG) experiment revealed the quantized nature of electron spins within neutral atoms~\cite{Gerlach1922}.  The same apparatus to create spin-polarized free electrons cannot be implemented due to particle momentum-position uncertainty and Lorentz force blurring~\cite{Mott1929}. Spin-polarized electron beams, which have become an indispensable tool for probing fundamental characteristics of matter ~\cite{Kuwahara2021,Pierce1980,Pierce1988,Kuwahara2014,Swartz1988}, have been produced through methods including spin-polarized photocathodes using semi-conductors~\cite{Pierce1975,Pierce1976,Pierce1980}, self-polarizing interactions of relativistic electrons in storage rings (the Sokolov-Ternov effect~\cite{Sokolov1964, Sokolov1967}), and proposed in high intensity free space interactions like the Kapitza-Dirac effect~\cite{Ahrens2012,Erhard2015, Wang2024}, or Compton-scattering~\cite{Bhatt1983}. These methods introduce challenges limiting their accessibility, including cathode maintenance for spin-polarized sources, and high laser intensities for free-space interactions. 

These challenges have promoted the continued exploration of methods to produce spin-polarized electrons. Work by Pan, et al.~\cite{Pan2023}, suggested the use of optical near fields to generate such beams. This method is attractive due to the reduced laser intensities required and the potential to achieve large beam polarization. Structure-mediated near-field-electron interactions have been demostrated in dielectric laser accelerators (DLAs) ~\cite{Peralta2013,Leedle2015,Leed20152,Leedle2018,Wootton2016, Breuer2013}, making similar geometries to implement spin-dependent interactions experimentally feasible.

Here, we propose a two stage approach using magnetic near-fields to produce spin-polarized electrons. We show an on-axis SG-type setup for free electrons to break spin-state symmetry~\cite{Batelaan1997,Batelaan2001}. We exploit intense magnetic field gradients created by rapid field oscillations of periodic grating near-fields, to impart efficient spin-dependent longitudinal forces parallel to the electron propagation axis ~\cite{Pan2023, Batelaan1997, Batelaan2001}. The second interaction stage polarizes electrons towards the same transverse spin-state. We show 70\% beam polarization, defined as the normalized difference between the number of electrons measured in one spin eigenstate and the other, can be achieved in the proposed devices, without loss of beam brightness or the need for external bending magnets. 

\begin{figure}[b]
    \centering
    \includegraphics[width=0.45\textwidth]{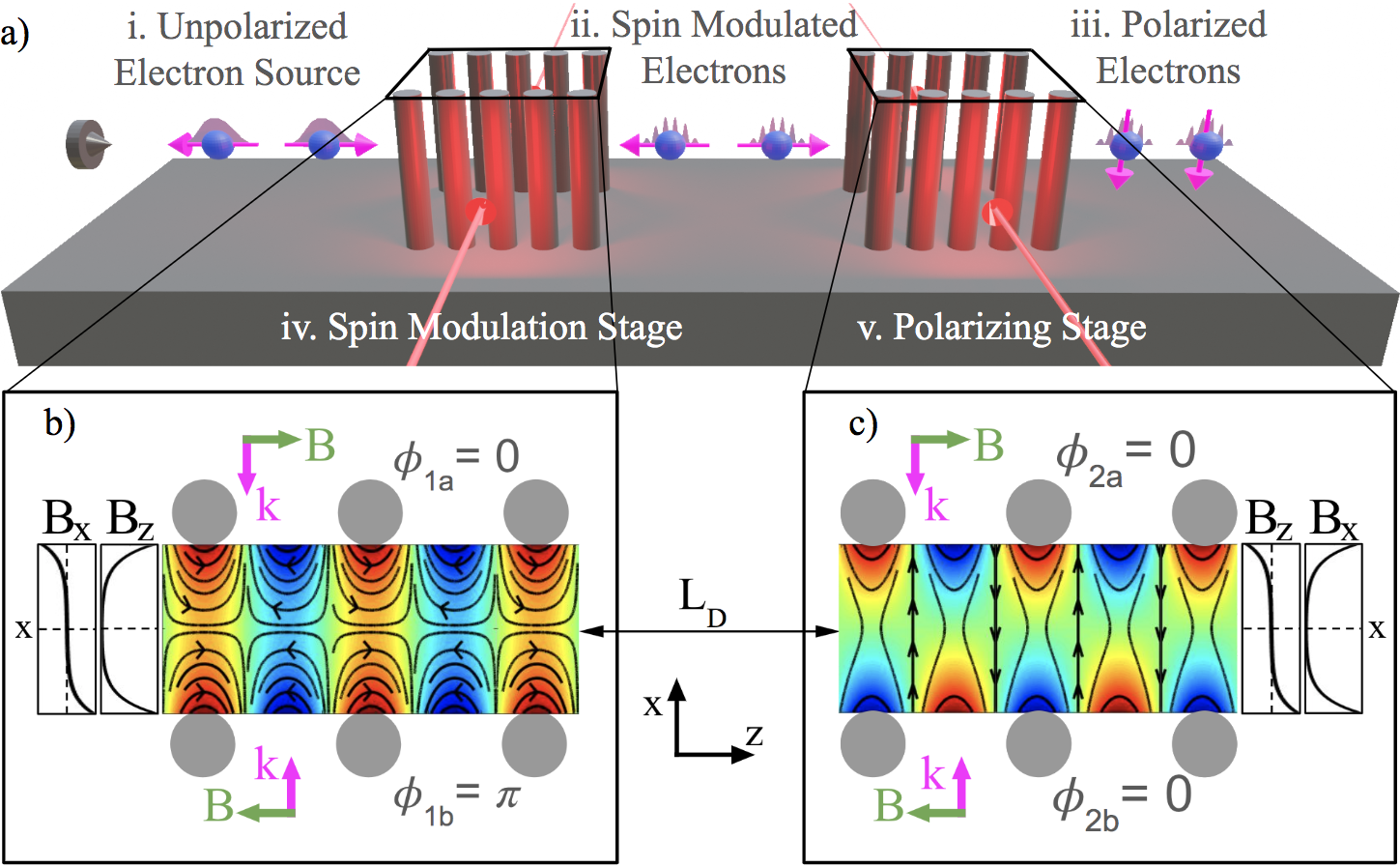} \caption{\textbf{Mechanism for generation of spin-polarized electron beams on a chip.} a.) Schematic of the proposed two-stage interaction set-up. The first stage and following free-space propagation introduces spin-dependent density formations and the second stage rotates the spin states to produce a polarized beam. b,c.) First harmonic magnetic field profiles of modulation and polarizing stages, respectively. Corresponding cosh and sinh transverse profiles of $B_x$ and $B_z$, and relative phases for defined illuminants are shown.}
    \label{fig:setup}
\end{figure}

\section{Spin Polarizer Operation}
The proposed spin-polarizing structure is shown in Figure 1, consisting of two stages of dual-pillars gratings separated by a drift length, $L_D$, where each pair of dual pillars along the gratings are separated by a channel width to allow electrons to propagate through the structure from left to right. Both dual pillar grating stages are illuminated on each side with intense counter-propagating transverse electric (TE) fields, where the magnetic field is polarized along the electron propagation axis, the $z$-axis. The illuminated sub-wavelength gratings produce channel modes that can be decomposed into near-field spatial harmonics ~\cite{Black2020}. Phase matching for synchronous propagation of the first harmonic and electrons traveling through the channel center, can be performed by tuning the laser wavelength and grating periodicity to fulfill $\lambda_g = \beta \lambda$, where $\beta = v_{el}/c$ is the electron's relativistic fraction, $\lambda$ is the free-space laser wavelength, and $\lambda_g$ is the grating period. The first-harmonic near-field oscillates sinusoidally along the electron propagation axis with the wavevector $k_z = 2\pi/\lambda_g$. A subrelativisitic electron phase-matched to near-infrared laser wavelengths has a $\lambda_g$ on the order of a few hundred nanometers, leading to rapid spatial oscillation of the near-field and a large amplitude magnetic field gradient force for spin-dependent interactions. 

When the dual pillar gratings are driven by counter-propagating plane waves, the evanescent near-fields interfere based on the relative phases between the plane waves. Figure 1b has the relative phases between the plane waves set so the longitudinal magnetic fields interfere constructively in the channel center, therefore creating what we denote as the first harmonic cosh mode for longitudinal spin-separation. The cosh mode exhibits cosh $B_z$ profile and a sinh $B_x$ profile along $\hat{x}$. Figure 1c has the relative phases between the plane waves set so the transverse magnetic fields interfere constructively in the channel center, therefore creating our denoted first harmonic sinh mode used for spin rotations. The sinh mode has a sinh $B_z$ profile and cosh $B_x$ profile along $\hat{x}$.

In this work we will consider the spin-dependent scatterings that occur in each stage of the polarizer by electrons with low-energy spreads, and consequentially large spatial spreads, along the propagation axis.  Low-energy spread electrons may span multiple periods of the sinusoidally oscillating spin-dependent force created in the spin-polarizer channel.  As observed in transverse magnetic (TM) photon-induced near-field electron microscopy (PINEM), where instead the magnetic field would be polarized along $\hat{y}$ ~\cite{Barwick2009,Park2010}, the sinusoidal force imparts energy modulation onto the electron wavepacket that leads to density bunch formations after a drift length~\cite{Park2010,Adiv2021,Baum2021}. In the next section we will formulate expressions for similar spin-dependent modulations occurring in the proposed structure when magnetic field interactions dominate due to TE illumination. 

\section{Theory}
\subsection{TE Spin-Density Bunch Formation}
To predict the spin-dependent bunching behavior of free electrons with unpolarized spins defined along the axis of propagation, $\hat{z}$, we consider a TE dual-drive field with the $\vec{E}$-field polarized along infinitely tall pillars (See Fig. 1a\b). The dual drive fields will be $\pi$ phase-shifted creating a cosh $B_z$ mode as shown in Fig. 1b~\cite{Black2020,Suppl}.

We consider an electron wave packet transversely defocused along the direction of the pillars, such that we can assume it is effectively diffractionless along this direction (see more details in SI Section V~\cite{Suppl}). The complex dimensionless magnetic coupling coefficient can be defined as, 
\begin{equation}
g_{B} = \frac{-i\mu_B k_z}{\hbar \omega } \int_{-L/2}^{L/2} B_z(z) e^{-ik_zz}dz,
\end{equation}
where  $\mu_B$ is the Bohr magneton,  $B_z(z)$ is the longitudinal magnetic field in the channel center, and $\omega$ is the laser angular frequency. With this definition we can perform time evolution on an initial electron wavepacket using the Pauli interaction Hamiltonian ~\cite{millerBook2009}, 
\begin{equation}
    \mathcal{\hat{H}}_I = \frac{e}{m_ec} \textbf{A}(\textbf{r},t) \cdot \hat{\textbf{p}} + \mu_b \textbf{B}\left(\textbf{r},t\right) \cdot \vec{\sigma} + \frac{e^2}{2m_ec^2} \textbf{A}^2\left(\textbf{r},t\right),
\end{equation}
where $\mathbf{A}(\mathbf{r},t)$ is the vector potential field, $\hat{\mathbf{p}}$ is the momentum operator, $m_e$ is the free electron mass, $e$ is electron charge, and $\vec{\sigma} = [\sigma_x,\sigma_y,\sigma_z]$ are the Pauli matrices. Unlike conventional PINEM with TM modes (where the electric field is aligned with the electron velocity), which is dominated by the $A\cdot p$ term in the Hamiltonian, here the coupling to the TE mode dominates through the $B\cdot\sigma$ term. Applying the time evolution operator defined from the interaction Hamiltonian and the Jacobi-Anger expansion, the wave function after the interaction is found. Following free-space drift, the quadratic photon-order--dependent accumulated phase can be included~\cite{Baum2021} to obtain the density-modulated wavefunction,
\begin{multline}
    \psi_\pm'(z') =\psi^0_\pm(z') \sum_n \mathcal{J}_n(2|g_{B_1}|) \exp\left(i n k_z z'+ in \phi_1 \right) \times\\
  \exp\left(in\left(-\frac{\pi}{2} \pm \frac{\pi}{2}\right)-in^2\pi\frac{L_D}{L_{QR}}\right),
\end{multline}
where $\psi^0_\pm(z')$ are the initial spin-dependent electron wavefunctions in space, $\mathcal{J}_n$ is the n$^\text{th}$ order Bessel function of the first kind, $g_{B_1}$ is the magnetic coupling coefficient defined in Eq. (1), $\phi_1$ is the argument of the magnetic coupling coefficient $g_{B_1}$, $z'$ is the position offset from the packet center $z_o$, $L_D$ is the free space drift length, and $L_{QR} = \beta^3 \gamma^3 m_e c \lambda^2/h$, is the quantum revival length as defined in~\cite{Baum2021}. The quantum revival length describes the free-space drift length the electron needs to travel in order to observe revivals of density distributions due to free electron's nonlinear dispersion. This result assumes a small initial longitudinal momentum spread, and negligible longitudinal dispersion over interactions lengths, which are much shorter than the quantum revival length.

\begin{figure*}[t]
    \centering
    \includegraphics[width=1\textwidth]{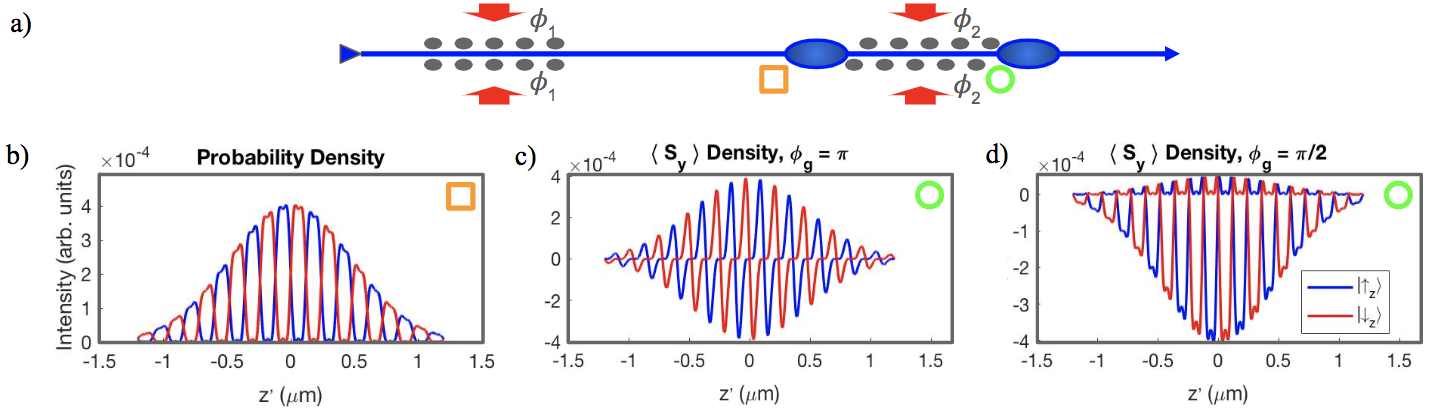} 
    \caption{\textbf{Spin-dependent electron densities.} a.) Experimental schematic. The positions where the electron density or spin density is calculated are highlighted by an orange square and a green circle. b.) Spin-dependent electron probability density distribution after interaction with first stage and free stage drift, input spin states differentiated by color. c,d) Electron y-axis spin density immediately following interaction with the second stage. c.) Has a relative interaction phase between stage 1 and 2 of $\pi$, leading to $\langle S_y \rangle \approx 0$. d.) has a relative interaction phase between stage 1 and 2 of $\pi/2$, which results in $\langle S_y \rangle \approx 0.7.$ }
    \label{fig:densities}
\end{figure*}

The $\pm$ subscript corresponds to electrons initially polarized in the spin-up or -down state along $\hat{z}$. The spin states are defined as $\ket{\uparrow} = [ \cos(\theta/2), \sin(\theta/2)e^{i\phi}]^T$ and $\ket{\downarrow} = [\sin(\theta/2),-\cos(\theta/2)e^{i\phi}]^T$, with polar coordinates $\theta = 0$ and $\phi = \pi$ used for the $\hat{z}$ spin quantization axis. The photon-order--dependent phase factors exhibit an additional  $\pi$ phase shift for odd photon-order interactions on spin-down electrons. The additional $\pi$ phase for the spin-down electrons leads to a spatial  $\pi$ phase-shift in density bunch formation, as seen in Fig. 2b. The probability distribution of initially $\ket{\uparrow_z}$ and $\ket{\downarrow_z}$ electron states is significantly separated into separate time bins following the interaction with the first stage and free space drift. There is minimal overlap between the up- and down-spin state's probability densities at $L_D =L_{QR}/2$, with small depolarizing contributions from the edges of the bunches; minimizing the overlap enhances the degree of achievable spin polarization. The sinh $B_x$ mode suppresses transverse field spin contributions along the channel center~\cite{Suppl}.

One of the main implications of Eq. (3) and the resulting time bin separation of Fig. 2b is the fact that one is able to map initial spin states $\ket{\uparrow_z}$ and $\ket{\downarrow_z}$, initially sharing the same spatial wavepacket $\psi^0$, into separate time bins, constituting a general unitary transformation $U_{\mathrm{SG}}$ that entangles the spin degree of freedom with longitudinal position. Explicitly, we have $U_{\mathrm{SG}}\ket{\uparrow_z,\psi^0}=\ket{\uparrow_z,\psi'_+}$ and $U_{\mathrm{SG}}\ket{\downarrow_z,\psi^0}=\ket{\downarrow_z,\psi'_-}$, where $\psi'_{\pm}$ are close to being orthogonal. Importantly, this transformation holds true also for initially mixed spin states $\rho_{\mathrm{spin}}\otimes \ket{\psi^0}\bra{\psi^0}$ (see further analysis in SI section VI). In the next section, we will demonstrate how the second interaction stage acts on the two wavepackets $\psi'_{\pm}$ to rotate their respective spins to be aligned with the same direction, thus effectively polarizing the beam.

\subsection{Creating High Polarization Electrons}
 In the previous section we detailed a method to create spin-dependent electron wavepacket distributions. These wavepackets can be used to create high polarization electron spin-distributions regardless of the initial spin state.

Achieving high polarization relies on the $\pi$ spatial phase shift between spin-dependent wavepackets, i.e. a $\ket{\uparrow_z}$ state is phase matched around $\phi_g$, while the $\ket{\downarrow_z}$ state is phased matched around $\phi_g+\pi$. Given a wave packet resulting from the spin-dependent modulation interacting with a cosh-transverse magnetic field mode (see Fig. 1c), the field will induce spin rotations. The direction and frequency of rotation will be interaction-phase dependent. Properly chosen interaction strengths, coupled with the spin-dependent spatial $\pi$ phase shifts in the probability density function, the expectation value of the final spin state can be polarized toward either $\ket{\uparrow_y}$ or $\ket{\downarrow_y}$ states along the y-axis. The dominating final spin-state can be chosen by tuning $\phi_g = \phi_1 - \phi_2$, the relative phase between the first and second interaction stages.

A similar coupling strength can be defined for the transverse field interaction of the second stage, by replacing $B_z(z)$ with $B_x(z)$ in Eq. (1). Following the second stage interaction, the state can be given by,
\color{black}
\begin{multline}
\ket{\psi''} =\int dz' \sum_{s=\uparrow,\downarrow} \psi_s'(z')\sum_m \mathcal{J}_m\left(2|g_{B_2}|\right) \times \\\exp\left(imk_zz'+im\phi_{2}\right) \sigma_x^m\ket{z',s_z},
\end{multline}
\color{black}
where $g_{B_2}$ and $\phi_{2}$ are the interaction strength and phase of the second stage, and $\ket{z',s_z} = \ket{z'} \otimes\ket{s_z}$ denotes the position and spin bases with eigenvalues $z'$ and $s_z \in \{\uparrow_z,\downarrow_z\}$ in which the wavefunction, $\psi'_s(z')$, as given in Eq. (3), is projected on. 

Expanding the summation in even-ordered, spin-preserving interactions, and odd-ordered, spin-flipping interactions, the spin-expectation value along the y-axis can be described for an arbitrary initial spin state,
\begin{multline}
    \langle \hat{S_y}\rangle =\\ \frac{\hbar}{2} \int dz' \left(|c_+|^2|\psi_+(z')|^2-|c_-|^2|\psi_-(z')|^2\right)\sin\left(2\alpha(z')\right)
    \\+2|c_+||c_-|\cos\left(2\alpha(z')\right) \Re\left\{ie^{i\phi_{d_\pm}}\psi_+(z')\psi_-^*(z')\right\}
\end{multline}
where $c_+$ and $c_-$ are the coefficients for an arbitrary initial superposition state of $\ket{\uparrow_z}$ and $\ket{\downarrow_z}$, respectively, $\phi_{d_\pm} = \arg(c_+)-\arg(c_-)$, and $\alpha(z') = 2|g_{B_2}|\sin(k_zz'+\phi_2)$. The pre-factor from Eq. (4) can be collapsed into the sine term in the spin-expectation value in Eq. (5). The outer sine term holds information about the cyclic nature of the spin flips. As the interaction becomes stronger (equivalently, as the $|g_{B_2}|$ factor increases), additional spin-flipping oscillations are introduced from the spatially varying field. This spatially varying spin factor is then multiplied onto the probability density function, to obtain a spin-weighted probability distribution. Eq. (5) can also describe the polarization of mixed spin states with density matrix $\rho_{\kappa\kappa'},~\kappa,\kappa'=\pm$, by replacing $|c_{\pm}|^2$ with $\rho_{\pm\pm}$ and $c_{\pm}c_{\mp}^*$ with $\rho_{\pm\mp}$.

Figure 2c and 2d show the effect of tuning the second stage interaction phase $\phi_{2}$ on the final spin distribution. Figure 2c shows an example where net spin polarization is not enhanced. Figure 2d, shows an example where the spin-polarization is enhanced and tuned towards a $\ket{\downarrow_y}$ state. The spin-weighted probability distribution shown in Fig. 2d equates to a spin expectation value of $\approx -0.7 \hbar/2$, which corresponds to a beam polarization percentage of approximately 70\%, or equivalently a probability of measuring a spin-down particle, $P_{\downarrow_y} \approx 85 \%$ and spin-up, $P_{\uparrow_y} \approx 15\%$. Calculations were performed considering initial spin states of $\ket{\uparrow_z}$ or $\ket{\downarrow_z}$, representing a maximally mixed input state $\rho = 1/2\ket{\uparrow_z}\bra{\uparrow_z} + 1/2\ket{\downarrow_z}\bra{\downarrow_z}$ when averaged. Thus, from the density matrix description and Eq. (5), various state mixtures resulting in a maximally mixed input beams can achieve $\approx 70\%$ polarization with our considered electron beam and laser parameters. These parameters include low energy spread electrons with $\sigma_E = 0.15 \;\text{eV}$, sub-relativistic initial electron energy with $\beta = 0.1$, and laser wavelength of 2.4 $\mu$m. Additionally, the first interaction region field strength  $E_y$ and interaction length $L$ is tuned so that only a couple orders of photon absorptions and emissions lead to significant contributions in the wave structure of the electron, or $|g_{B_1}| \approx 0.5$. After the spin-dependent modulation of the first stage, a drift length $L_{QR}/2$ is chosen for maximal contrast~\cite{Baum2021}. The second interaction length and field strength is tuned so $|g_{B_2}| \approx 0.5$, to maximize the spin-flipping first-order photon exchange and minimize higher order exchanges.

The discussed set of conditions give a large overlap between portions of high $z$-axis spin density and spin-flipping fields, enabling the creation of highly polarized electrons. The use of moderate $|g_B|$ factors benefits coherence lengths for density bunches so that quadratic phase accumulation within interaction regions can be reasonably neglected. The sub-wavepacket spin separation inherent to the proposed interaction, enables the use of longitudinally extended electrons further improving coherence lengths.  

Figure 3 maps the value of $\langle \hat{S}_y \rangle$ when $\phi_g = \pi/2$ in two cases. Figure 3a maps varying $|g_{B_{1/2}}|$ and $L_D/L_{QR}$, when $|g_{B_1}| = |g_{B_2}|$. High spin-polarization effects are observed over a significant range of $L_D$ values for moderate magnetic coupling coefficients, while increased magnetic coupling factors see a reduction in spin polarization magnitude and acceptable drift range. Figure 3b depicts the one-to-one linear relationship between the coupling factor in the first and second stage when $L_D$ is half $L_{QR}$. The moderate and equal coupling coefficients exhibit significant spin-polarization and strong temporal coherence.

\begin{figure}[t]
    \centering
    \includegraphics[width=0.45\textwidth]{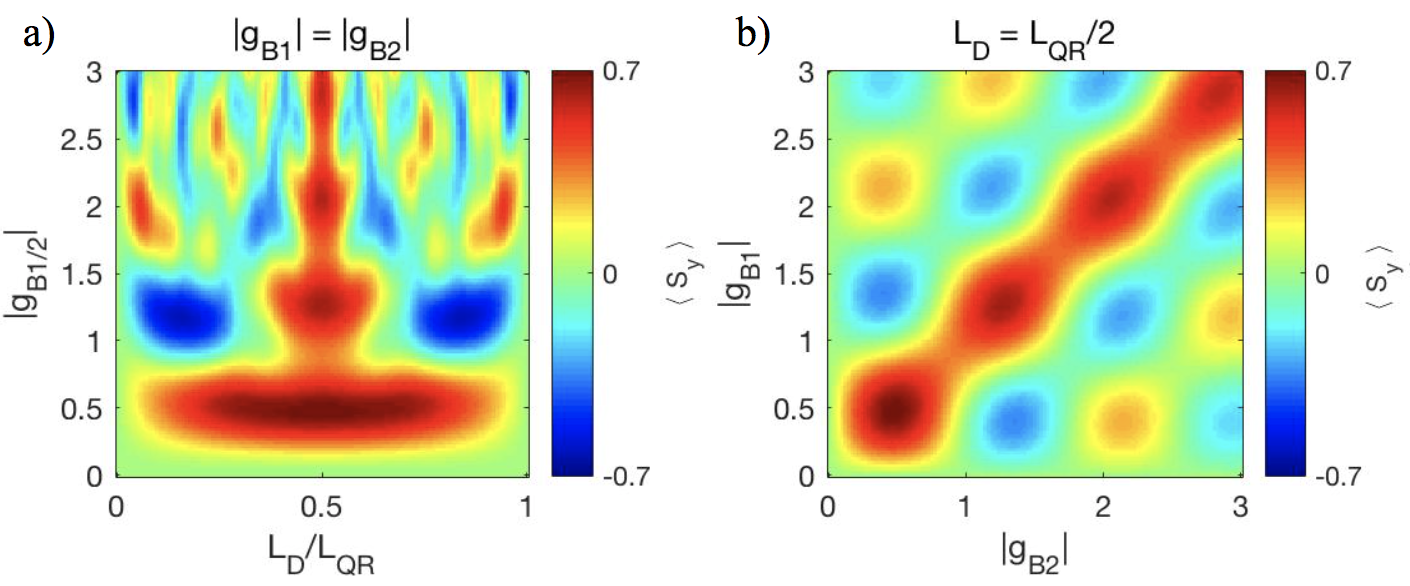}
    \caption{\textbf{Influence of experimental parameters.}  a) Heat map of $\langle S_y \rangle$ in terms of parameters drift length and coupling strength, with the condition $|g_{B1}| = |g_{B2}|$. Notably, high polarization shows minimized drift length dependence for $|g_{B_1}| = |g_{B_2}| \approx 0.5$ corresponding to microbunches with significant temporal coherence. b) Heat map of $\langle S_y \rangle$  in terms of $|g_{B1}|$ and $|g_{B2}|$, depicting higher order overlaps when the drift length is half the quantum revival length.}
    \label{fig:params}
\end{figure}

\section{Physical Considerations}
A commercial femtosecond laser can provide the needed fields for spin-polarization on-a-chip. Considering the requirement of $|g_{B_{1/2}}| \approx 0.5$ and the electron beam parameters discussed above, we choose an interaction length of 12.7 $\mu$m for each interaction stage, an achievable conversion efficiency from incident field strength to field strength in the channel center of 0.35, and incident electric field of 550 MV/m  on each dual drive side. Fields of 550 MV/m have been shown to be under the laser induced damage threshold (LIDT) of hydrogen-annealed processed silicon pillars for 300 fs, 1.96 $\mu$m laser pulses~\cite{Miao2020}.   Longer wavelengths are expected to exhibit higher LIDTs~\cite{Werner2019}. The use of a pulse front tilt (PFT) has been demonstrated to improve interaction time~\cite{Cesar18, Broaddus2024}, and could be used with shorter laser pulses to further increase peak fields, reducing required interaction lengths.

Realisitic electron beam parameters are important for experimental implementation. We note, due to the only electric field contributions being $E_y$, the $\hat{p} \cdot \hat{A}$ term of the Pauli Hamiltonian that was neglected previously would be sensitive to $v_y$ within the transverse electron wavepacket. Additionally, for the first stage where PINEM interactions would degrade the fidelity of the spin-dependent microbunches, $E_y$ exhibits a sinh mode along the $x$-axis around the center of the channel, and thus vanishes at the position of the electron beam. 

 In the case that the electron beam is highly collimated in the y-direction and focused in the $x$-direction throughout the interaction region, non-zero photon order terms can be heavily attenuated. An astigmatic beam with such characteristics can provide high fidelity spin distributions and naturally fits the geometry of a DLA with a narrow $x$-axis channel, and very tall pillars in the $y$-direction (see Suppl Information Sect. V ~\cite{Suppl}).

\section{Conclusions}
We have proposed a single-chip, two-stage,  device to produce a polarized electron beam from an initially unpolarized beam. By converting spin-orthogonalities into time-based orthogonalities, the two stage interaction is a powerful operation to polarize arbitrary unpolarized initial state mixtures.  No external bending magnets are required for the creation of high polarization beams. Switching between final state polarization can be performed by tuning the relative phases of the two stages. Reducing the requirement of external magnetic fields opens the door for efficient on-axis integration with PINEM and other photonic on-chip technology, such as focusing structures or photonic guiding structures. Our proposed method has theoretical limits above that of a standard GaAsP spin-polarized cathode~\cite{Grames2023}, which requires intense upkeep. Other methods to spin-polarize electrons have produced higher polarizations than presented ~\cite{Grames2023}, but introduce additional challenges in upkeep and technology access to large-scale facilities and equipment. 

 Beyond creation of significantly polarized electrons, near-field based momentum modulations and spin-rotations can create more complex spin superposition states and probability-density structures within the electron wave packet. The wavepacket formalism is of interest to free-electron quantum optics, an increasingly explored research area~\cite{Ruimy2025}, where spin-dependent dynamics may add interesting additional degrees of freedom to carry and manipulate quantum information held by free-electrons. The fine spin and momentum structure created within the electron wave packet could present TE and TM PINEM prepared electrons as unique probes for ultrafast experiments and imaging of spin-dependent properties.

\bibliographystyle{apsrev}
\bibliography{references}

\end{document}

\title{On-Chip Laser-Driven Free-Electron Spin Polarizer Supplementary Information}

\author{Clarisse Woodahl\thanks{Corresponding author: cwoodahl@stanford.edu}}
\email{cwoodahl@stanford.com}
\author{Melanie Murillo}
\author{Charles Roques-Carmes}
\author{Aviv Karnieli}
\author{David A. B. Miller}
\author{Olav Solgaard}
\affiliation{E. L. Ginzton Laboratory, Stanford University, Stanford, CA, 94305, U.S.A.}

\date{\today}

\maketitle

\section{Contents}
I. Dual Drive Field Modes

II. Spin-Dependent Density Bunch Formalism

III. Spin-Flip formalism

IV. Y-Spin State Expectation Derivation

V. Lorentz Considerations

VI. Arbitrary Initial Spin States

\section{I. Dual Drive Field Modes}
Considering the geometry presented in main text Fig. 1, the fields created by the dual drive symmetric infinitely tall pillars can be idealized to impart no Lorentz forces. We consider phase-matching to the first spatial harmonic of the near-fields. First-order harmonics have the slowest decay length enabling larger channel widths and beam acceptances.

In this consideration we can write the transverse electric (TE) driven electric field of the first-harmonic as a detailed in ~\cite{Black2020},
\[\vec{E} (x,z) = \left[0,\;E_y \left(b_1^+ \cosh\left(\Gamma x\right)+b^-_1 \sinh\left(\Gamma x\right)\right)e^{-ik_gz},\;0\right]\]

where there are contributions to the electric field only along the y-axis. $\Gamma$ is the mode decay length of the first harmonic. As defined in ~\cite{Black2020}, the complex coefficients for TE modes are,
\[b^\pm_n = (1 \pm r_n) \frac{e^{i\theta_r} \pm 1}{2}\]

Where $r_n$ is the complex ratio of a mode created by each dual-drive illuminant, and $\theta_r$ is the relative phase between the two illuminants. The magnetic field can then be written, 
\begin{multline*}
    \vec{B}(x,z) = \frac{iE_y k_z}{\omega} e^{ik_zz} \times 
    \\
    \left[b_1^+\cosh(\Gamma x)+b_1^- \sinh(\Gamma x), \;0, 
    \; b_1^+ \sinh(\Gamma x)+b_1^- \cosh(\Gamma x)\right]
\end{multline*}

As discussed in \cite{Black2020}, for a transversely diffractionless electron wave packet, there would be no Lorentz forces. In the case where $r_n \approx 0$, and $\theta_n = \pi$, as in Fig. 1a creates sinh modes in $E_y$ and $B_x$, and a cosh mode in $B_z$. Along the center of the channel the field then has $E_y = 0$, $B_x = 0$ and $B_z \neq 0$. This geometry minimizes transverse magnetic field components, reducing interactions with spin-flipping fields. The geometry produces the spin-dependent on-axis force leading to spin-dependent  density bunch formation. The force amplitude oscillates with $e^{ik_zz}$, providing position dependent accelerating or decelerating force.

For Fig. 1b, changing $\theta_r = 0$, creates a sinh mode in $B_z$, and cosh modes in $E_y$ and $B_x$. In the center of the channel the spin flipping field $B_x \neq 0$, while $B_z = 0$. In this geometry spin-flip interactions can dominate. The field amplitude oscillates with $e^{ik_zz}$, providing a position or phase dependent spin-flip strength and direction.

\section{II. Spin-Dependent Density Bunch Formalism}

We consider the Pauli-Hamiltonian given by Eq. (2) of the main text. As done in ~\cite{Pan2023}, with idealized electron beam and field parameters, the interaction Hamiltonian can be simplified to $\mathcal{\hat{H}}_I = \mu_B \hat{B} \cdot \sigma$. This term holds the spin-dependent electron-light interactions in the sub-relativistic regime. The time evolution operator can written as, 
\[U(t,t_o) = \exp\left(-\frac{i}{\hbar}\int dt' \mu_B \vec{B} \cdot \vec{\sigma} \right)\]

In the radiation gauge the magnetic field can be given as $\vec{B} = \vec{\nabla} \times \vec{A}$, where $\vec{A}$ is the magnetic potential vector field $\vec{E} = -\partial /\partial t \vec{A} (r,t)$.  We can write the electric field $\vec{E} (\vec{r},t) = \vec{E}(\vec{r})e^{-i \omega t} + c.c.$ Expanding the magnetic field for the cosh dual drive mode, where along the center of the channel there is only non-zero longitudinal magnetic field that sinusoidally oscillates, we can rewrite the time evolution operator as,
\begin{multline*}
U(t,t_o) = \\\exp\left(-\frac{i\mu_B}{\hbar }\left[
\begin{matrix}
1 & 0 \\
0 & -1
\end{matrix}
\right]
\int dt' B_z(\vec{r})e^{-i \omega t'} + B_z^*(\vec{r})e^{i \omega t'}\right)
\end{multline*}

Additionally, in the low $\beta$ regime the transverse wavevector $k_x$ and longintudinal wavevector $k_z$ of the near-field mode are approximately equal. Replacing $t'$ with $(z-z')/v_e$ ~\cite{Baum2021}, considering the central trajectory along the channel, and noting $\omega/v_e = k_z$ the time evolution operator can be then written
\begin{multline*}
    U(t,t_o) = \\
    \exp\left(-\frac{i\mu_B k_z}{\hbar \omega} \left[
    \begin{matrix}
    1 & 0 \\
    0 & -1
    \end{matrix}
    \right]2 \Re \left\{ e^{ik_z z'} \int dz B_z(z)e^{-i k_z z}\right\}\right)
\end{multline*}

Recalling the definition from the main text of the complex magnetic coupling constant, 
\[g_B = \frac{-i\mu_B k_z}{\hbar \omega} \int_{-L/2}^{L/2} B_z(z) e^{-ik_zz}dz\]

The time evolution operator can be further simplified,
\[ U(t,t_o) = \exp\left(\left[
\begin{matrix}
1 & 0 \\
0 & -1
\end{matrix}\right]2 |g_B| i \sin(\phi_1 + k_zz')\right)\]

Where $\phi_1 = \arg(g_B)$. Then using the Jacobi-Anger expansion, the time evolution exponential operation can be written in terms of a bessel summation,
\begin{multline*} U(t,t_o) =\\ \sum_n J_n(2|g_B|) \exp\left(ink_zz'+in\phi_1 + in \left(\frac{\pi}{2} \mp \frac{\pi}{2}\right)\right)
\end{multline*}
 The $\sigma_z$ matrix in the exponential can be collapsed into the final term of the phase exponential when expanding the exponential using the Jacobi-Anger expansion and bessel expansion. Electrons initially in the $|\uparrow_z\rangle$ state see no additional spin dependent phase shift from the Pauli-z matrix, while spins in the $|\downarrow_z\rangle$  state see an additional $e^{in\pi}$ phase factor from the $\sigma_z$ matrix.

The spin-dependent phase factor introduced onto odd-ordered interactions create interferences patterns that correspond to the $\pi$ spatial phase shift between $|\uparrow_z\rangle$ and $|\downarrow_z\rangle$ states after free-space propagation. 

Thus far we considered the initial state to be either a $|\uparrow_z\rangle$ or $|\downarrow_z\rangle$ state. See SI section VI. for a discussion of arbitrary mixed states, and the Stern-Gerlach like separation that occurs longitudinally when arbitrary superposition states are considered.

\section{III. Spin Flip Formalism}
Performing a similar simplification of the time evolution operator as in SI I., for a sinh longitudinal magnetic field, or a cosh magnetic field in the x-direction we obtain the following expression for the time-evolution operator.

\[U(t,t_o) = \sum_m J_m\left(2|g_B|\right) \exp\left(imk_zz'+im\phi_2\right) \left[
\begin{matrix}
0 & 1\\
1 & 0\end{matrix}\right]^m\]

Even order photon interaction correspond to a matrix factor $\sigma_x^{2m} = \mathbf{I}^m = \mathbf{I}$, which perform spin preserving interactions. Odd-order photon interactions correspond to a matrix factor  $\sigma_x^{2m+1} = \mathbf{I} \sigma_x = \sigma_x$ , which induces spin flips.

We are now in a position to investigate how the entire interaction operates on an initial electron beam spin state. Considering an electron wave function described as a spatial distribution and a spin state $\ket{\uparrow_z}$ given as,

\begin{multline*}
\ket{\psi'_+} =\int dz' \psi'_+(z') \ket{z'}\ket{\uparrow_z} \\=  \int dz' \psi'_+(z') \ket{z'}\frac{\ket{\uparrow_y}+\ket{\downarrow_y}}{\sqrt{2}}  
\end{multline*}
\color{black}
And similarly for $\ket{\downarrow_z}$,
\begin{multline*}
\ket{\psi'_-} =\int dz' \psi'_-(z') \ket{z'}\ket{\downarrow_z} \\=  -\int dz' \psi'_-(z') \ket{z'}i\frac{\ket{\uparrow_y}-\ket{\downarrow_y}}{\sqrt{2}}  
\end{multline*}
\color{black}

where the states $| \uparrow_y \rangle$ and $| \downarrow_y \rangle$ are up- and down-states along the y-axis. We choose to write the z-axis spin state as a superposition of y-axis states because we are interested in spin expectation values along the y-axis.

Considering the Bloch sphere representation the z-axis states can be written as a superposition of y-axis states: $|\uparrow_z\rangle = \left(|\uparrow_y\rangle + |\downarrow_y\rangle\right)/\sqrt{2}$, and $|\downarrow_z\rangle = -i\left(|\uparrow_y\rangle - |\downarrow_y\rangle\right)/\sqrt{2}$. In doing so we notice for the $|\uparrow_z\rangle$ state the y-spin states are in phase, whereas for the $|\downarrow_z\rangle$ state the y-spin states are  phase shifted to each other.

We can consider the odd and even ordered photon interactions breaking the time evolution operator into two parts,
\[\begin{matrix}
U(t,t_o) = \sum_m J_{2m}(2|g_B|)e^{i2mk_zz'+i2m\phi_2}

\left[
\begin{matrix}
1 & 0 \\
0 & 1
\end{matrix}
\right]
+ \\

\;\;\;\;\;\;\;\;\;\;\;\;\;\;\;\sum_m J_{2m-1}(2|g_B|) e^{i(2m-1)k_zz'+i(2m-1)\phi_2}
\left[
\begin{matrix}
0 & 1 \\
1 & 0
\end{matrix}
\right]
\end{matrix}
\]

Using the property of Bessel functions that $J_{2m}(x) = J_{-2m}(x)$ and $J_{2m-1}(x) = -J_{-(2m-1)}(x)$ we can rewrite the Bessel,
\begin{multline*}
    U(t,t_o) =\\\left(J_o(2|g_B|)+
    2\sum_{m=1}^\infty J_{2m}(2|g_B|)\cos(2m(k_zz'+\phi_g))\right) \mathbf{I} +\\
    2i\sum_{m=1}^\infty J_{2m-1}(2|g_B|)\sin((2m-1)(k_zz'+\phi_g))\left[\begin{matrix} 0 & 1 \\ 1 & 0 \end{matrix}\right]
\end{multline*}
which, using the real-valued Jacobi-Anger expansions, can be simplified to
\begin{multline*}
U(t,t_o)= \cos\left(2|g_B|\sin(k_zz'+\phi_2)\right)\left[\begin{matrix}1 & 0 \\ 0 & 1\end{matrix}\right] + \\ i\sin\left(2|g_B|\sin(k_zz'+\phi_2)\right)\left[\begin{matrix}0 & 1 \\ 1& 0\end{matrix}\right]
\end{multline*}

Operating the time-evolution operator on spin-states as discussed before results in a new electron wavefunction, 
\begin{multline*} 
\ket{\psi''} = \int dz'\psi'(z') \ket{z'}\times \left( \left(c^\uparrow \cos(\alpha(z')) + c^\downarrow\sin(\alpha(z'))\right)\ket{\uparrow_y} \right.\\\left.+ \left(c^\downarrow \cos(\alpha(z'))-c^\uparrow \sin(\alpha(z'))\right)\ket{\downarrow_y}\right)
\end{multline*}

With specific states given as a combination of spin-$z$ and spin-$y$ states as, ,

\color{black}
\begin{multline*}
\ket{\psi''_+} =\int dz'\psi'_+(z')\ket{z'}\left[\cos\alpha(z')  \ket{\uparrow_z} +i \sin\alpha(z') \ket{\downarrow_z} \right]\\=   \int dz'\psi'_+(z')\ket{z'}\Bigg[\cos\alpha(z')  \frac{\ket{\uparrow_y}+\ket{\downarrow_y}}{\sqrt{2}}\\ + \sin\alpha(z') \frac{\ket{\uparrow_y}-\ket{\downarrow_y}}{\sqrt{2}} \Bigg] \\ =   \int dz'\psi'_+(z')\ket{z'}\Bigg[\sin \left(\frac{\pi}{4}+\alpha(z')\right)  \ket{\uparrow_y}\\ + \sin \left(\frac{\pi}{4}-\alpha(z')\right)\ket{\downarrow_y} \Bigg]
\end{multline*}
\color{black}

\color{black}
\begin{multline*}
\ket{\psi''_-} =\int dz'\psi'_-(z')\ket{z'}\left[\cos\alpha(z')  \ket{\downarrow_z} +i \sin\alpha(z') \ket{\uparrow_z} \right]\\=  i \int dz'\psi'_-(z')\ket{z'}\Bigg[-\cos\alpha(z')  \frac{\ket{\uparrow_y}-\ket{\downarrow_y}}{\sqrt{2}}\\ + \sin\alpha(z') \frac{\ket{\uparrow_y}+\ket{\downarrow_y}}{\sqrt{2}} \Bigg] \\ =  i \int dz'\psi'_-(z')\ket{z'}\Bigg[-\sin \left(\frac{\pi}{4}-\alpha(z')\right)  \ket{\uparrow_y}\\ + \sin \left(\frac{\pi}{4}+\alpha(z')\right)\ket{\downarrow_y} \Bigg]
\end{multline*}
\color{black}
where $\alpha(z') = 2|g_B|\sin(k_zz'+\phi_2)$.

\section{IV. Expectation Value Derivation}
Using the electron wave function derived in the previous section, we can evaluate the spin expectation value. 
\begin{multline}
    \langle \hat{S_y}\rangle =\\ \frac{\hbar}{2} \int dz' \left(|c_+|^2|\psi_+(z')|^2-|c_-|^2|\psi_-(z')|^2\right)\sin\left(2\alpha(z')\right)
    \\+2|c_+||c_-|\cos\left(2\alpha(z')\right) \Re\left\{ie^{i\phi_{d_\pm}}\psi_+(z')\psi_-^*(z')\right\}
\end{multline}

Where $\phi_{d_\pm} = \phi^+ - \phi^-$, when $c^+ = |c^+|e^{i\phi^+}$ and similar for the down state. If we consider the case where the initial state is defined either up or down along the z-axis, then $\{|c^+|,|c^-|\} = \{1,0\}$ or $\{0,1\}$ 
\[
\langle \psi''_\pm|\hat{S}_y|\psi''_\pm\rangle = \frac{\hbar}{2}\int dz' |\psi'_\pm(z')|^2 \left(\pm\sin(2\alpha)\right)
\]
The nested sine term holds to cyclic nature of the spin flip rotations, accounting for the spatial variation in spin-rotating  fields. 

\color{black}
In addition to expectation values, one can also calculate the probability for measuring either $\uparrow_y$ or $\downarrow_y$:

\begin{multline*}
    P_{\uparrow,\pm} = |\braket{\uparrow_y|\psi''_{\pm}}|^2 = \int dz' |\psi'_{\pm} (z')|^2 \sin^2 \left[\alpha(z')\pm \frac{\pi}{4}\right]
\end{multline*}
\begin{multline*}
    P_{\downarrow,\pm}=|\braket{\downarrow_y|\psi''_{\pm}}|^2 = \int dz' |\psi'_{\pm} (z')|^2 \sin^2 \left[\alpha(z')\mp \frac{\pi}{4}\right]
\end{multline*}
From these expressions, it is clear how the overlap of the different modulated wavefunctions with the shifted sine squared term determines the probability for measuring a definite value for the spin component along y. Choosing the interaction parameters for optimizing the shape of $\psi_{\pm}(z')$ and $\alpha(z')$ to maximize, say, $P_{\uparrow,\pm}$ (which dictates minimizing $P_{\downarrow,\pm}$) is the key to our proposal. For a general initial state in the $z$ basis written as $a_+\ket{\uparrow_z}+a_-\ket{\downarrow_z}$, if we approximate $\psi_+(z')\psi_-(z')\approx 0$ owing to the initial SG separation into different time bins, cross-terms approximately do not contribute and we have
\begin{multline*}
    P_{\uparrow} = |a_+|^2\int dz' |\psi'_+ (z')|^2 \sin^2 \left[\alpha(z')+ \frac{\pi}{4}\right] \\ + |a_-|^2\int dz' |\psi'_- (z')|^2 \sin^2 \left[\alpha(z')- \frac{\pi}{4}\right]
\end{multline*}
\begin{multline*}
    P_{\downarrow} = |a_+|^2\int dz' |\psi'_+ (z')|^2 \sin^2 \left[\alpha(z')- \frac{\pi}{4}\right] \\ + |a_-|^2\int dz' |\psi'_- (z')|^2 \sin^2 \left[\alpha(z')+ \frac{\pi}{4}\right]
\end{multline*}
Again, one aims to maximize, say $P_{\uparrow}$ (minimize $P_{\downarrow}$). Finally, since cross-terms approximately do not contribute, we can replace the initial pure state with an initial mixed state $\rho$, for which $|a_{\pm}|^2$ above can be replaced by $\rho_{++}$ and $\rho_{--}$, respectively (see also Section SI VI).

\color{black}

\section{V. Lorentz Considerations}
In realistic experiments the magnitude of Lorentz forces from diffracting beams is important to consider. Consider a three dimensional separable electron wave packet, $\psi(\vec{r}) = \psi_x(x)\psi_y(y)\psi_z(z)$. Due to the separability we can express $\psi_y(y)$ as the Fourier transform, as $\tilde{\psi_y}(p_y)$. 

Considering the Pauli Hamiltonian term  as the interaction Hamiltonian, $\hat{H}_I = q/m_e \; \left(\hat{\mathbf{p}} \cdot \mathbf{A}\right)$. Due to only $E_y$ there is only $A_y$ contributions, the interaction Hamiltonian becomes, $\hat{H}_I = q/m_e \; \left(A_y(x,z) \; \hat{p}_y\right)$. The time evolution operator can then be written,
\begin{multline*}
U(t,t_o) = \\\exp\left(\frac{-q}{\hbar \omega m_e} \int \left(E_y(x,z)e^{i\omega t} - E^*_y(x,z)e^{-i \omega t}\right) \hat{p}_y dt\right)
\end{multline*}

Which can be written using the Lorentz complex coupling coefficient,
\[g_L(x) = -\frac{q}{\hbar \omega} \int E_y(x,z)e^{-ik_zz} dz\]

The time evolution operator can then be written,
\[U(t,t_o) = exp\left(\frac{2i |g_L(x)|}{ m_ev_{e}} \sin\left(k_zz'+\phi_{g_L}(x)\right) \hat{p}_y \right)\]
 The time evolution operator can then be expanded as a summation into,
\[U(t,t_o) = \sum_n \frac{1}{n!} \times \left(\frac{2 i|g_L(x)|}{m_ev_{e}} \sin \left(k_zz'+\phi_{g_L}(x)\right)\right)^n (\hat{p_y})^n\]

When considering such a time evolution operator, operating on the  separable electron wavefunction defined previously results in $(\hat{p}_y)^n \tilde{\psi}_y(p_y) = (p_y)^n \tilde{\psi}_y(p_y)$, the time evolution operator can be written then,
\[U(t,t_o) = \exp\left(2 i |g_L(x)|\frac{p_y}{ m_ev_{e}} \sin\left(k_zz'+\phi_{g_L}(x)\right) \right)\]

Then using the Jacobi-Anger expansion, the time evolution operator can be rewritten,
\[U(t,t_o) = \sum_n J_n\left(2|g_L(x)|\frac{p_y}{m_ev_{e}}\right)\exp\left(in(k_zz'+\phi_{g_L}(x))\right)\]

\begin{figure}[h]
    \centering
    \includegraphics[width=0.49\textwidth]{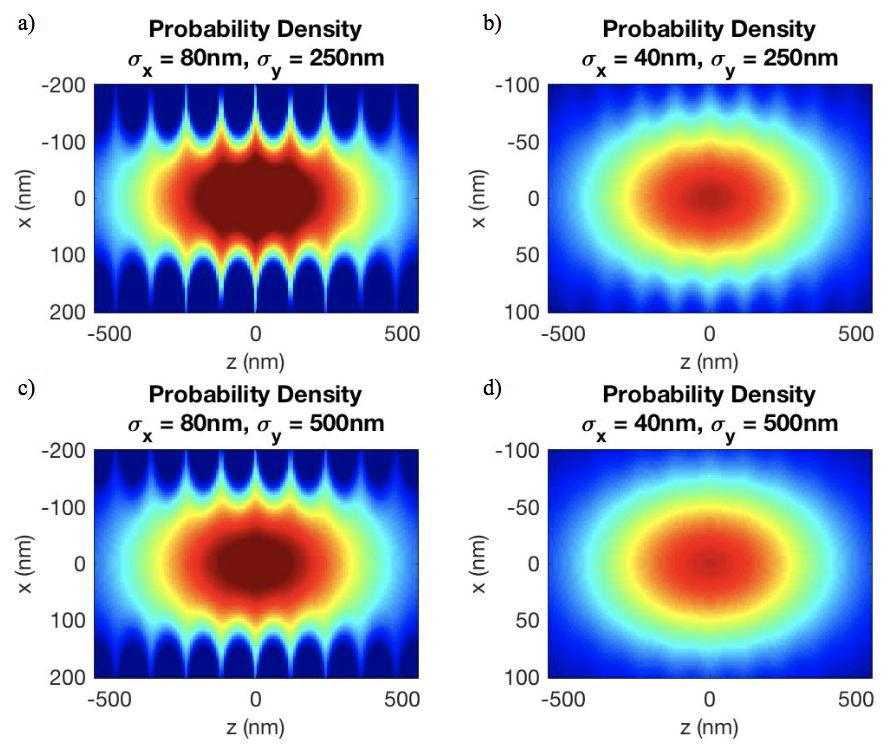}
    \caption{Probability density function of an electron considering only PINEM interactions with parameters discussed in the main text for different x-axis and y-axis position uncertainties. Under the assumption the interaction region is sufficiently short that diffraction is minimal during the interaction region. a.) exhibits the largest PINEM bunched probability density due to its large $\Delta p_y$ and large $\Delta x$, b,c) intermediate steps of improving resistance to PINEM scatterings by reducing $\Delta x$ (b) or decreasing $\Delta p_y$ (c), d.) combination of both changes from (b) and (c) exhibiting little PINEM structures within the wavepacket.}
    \label{fig:setup}
\end{figure}

Leading to the equation,
\begin{multline*}
    \psi'_{PINEM}(x,z) = \int dp_y \sum_n \:J_n \left(2|g_L(x)|\frac{p_y}{m_ev_e}\right) \\
    \times \; \; e^{in(k_zz'+\phi_{g_E}(x))-in^2\pi L/L_{QR}} \psi_x(x) \tilde{\psi}_y(p_y)\psi_z(z)
\end{multline*}

Considering the sinh $E_y$ mode along the x-axis of the first stage, as well as a gaussian $\psi_x(x)$, the overlap can be minimized for the interaction length. Additionally, a largely collimated electron beam along the y-axis, can further reduce the PINEM interaction strength.

We can predict the intensity of the PINEM interaction taking into account reasonable electron beam parameters. An astigmatism in the electron beam enables optimal PINEM suppression. Considering a separable, gaussian wave packet in each direction, with specified $\sigma_x$ and $\sigma_y$ and the beam and laser parameters discussed in the main text we can predict the PINEM interaction strength.

Using the real-valued Jacobi-Anger expansions, the symmetric nature of the momentum wave function $\tilde{\psi}_y(p_y)$ we consider, and the quadratic-order dependent phase accumulation at half the Talbot revival length the wave function can be further simplified,
\begin{multline*}
\psi'_{PINEM}(x,z) = \int dp_y \times\\ \cos\left(2|g_L(x)|\frac{p_y}{m_ev_e}\sin(k_zz'+\phi_{g_L}(x))\right)\times\\\psi_x(x) \tilde{\psi}_y(p_y)\psi_z(z)
\end{multline*}

\section{VI. Arbitrary Spin State Considerations}
In the main text we quantized the initial spin states along the z-axis. It is of interest to consider other arbitrary initial electron spins.

An electron beam with a zero polarization vector, with no preferential polarization along any axis, has a density matrix of the identity matrix divided by two ~\cite{bransden2000quantum}. Ensemble average values remain consistent regardless of quantization or representation of an unpolarized beam due to the equivalent density matrix and the ensemble average, $\left[A\right] = Tr\{\rho A\}$.

The first of the two stages separates spin density within the electron wave packet. In the main text we considered only electrons with one of two longitudinal spins values, here we consider initial spin states of an unpolarized electron beam that are linear superpositions of z-axis up and down spin. Corresponding to x-axis up and down spin, 
\[|\uparrow_x\rangle = 1/\sqrt{2} |\uparrow_z\rangle + 1/\sqrt{2} |\downarrow_z \rangle\]
\[|\downarrow _x\rangle = 1/\sqrt{2} |\uparrow_z \rangle - 1/\sqrt{2} |\downarrow_z \rangle\]
and to y-axis up and down spin,
\[|\uparrow_y\rangle = 1/\sqrt{2} |\uparrow_z\rangle + i/\sqrt{2} |\downarrow_z \rangle\]
\[|\uparrow_y\rangle = 1/\sqrt{2} |\uparrow_z\rangle - i/\sqrt{2} |\downarrow_z \rangle\]

The z-axis spin density after half the Talbot revival length is shown in Figure S2.

\begin{figure}[h]
    \centering
    \includegraphics[width=0.49\textwidth]{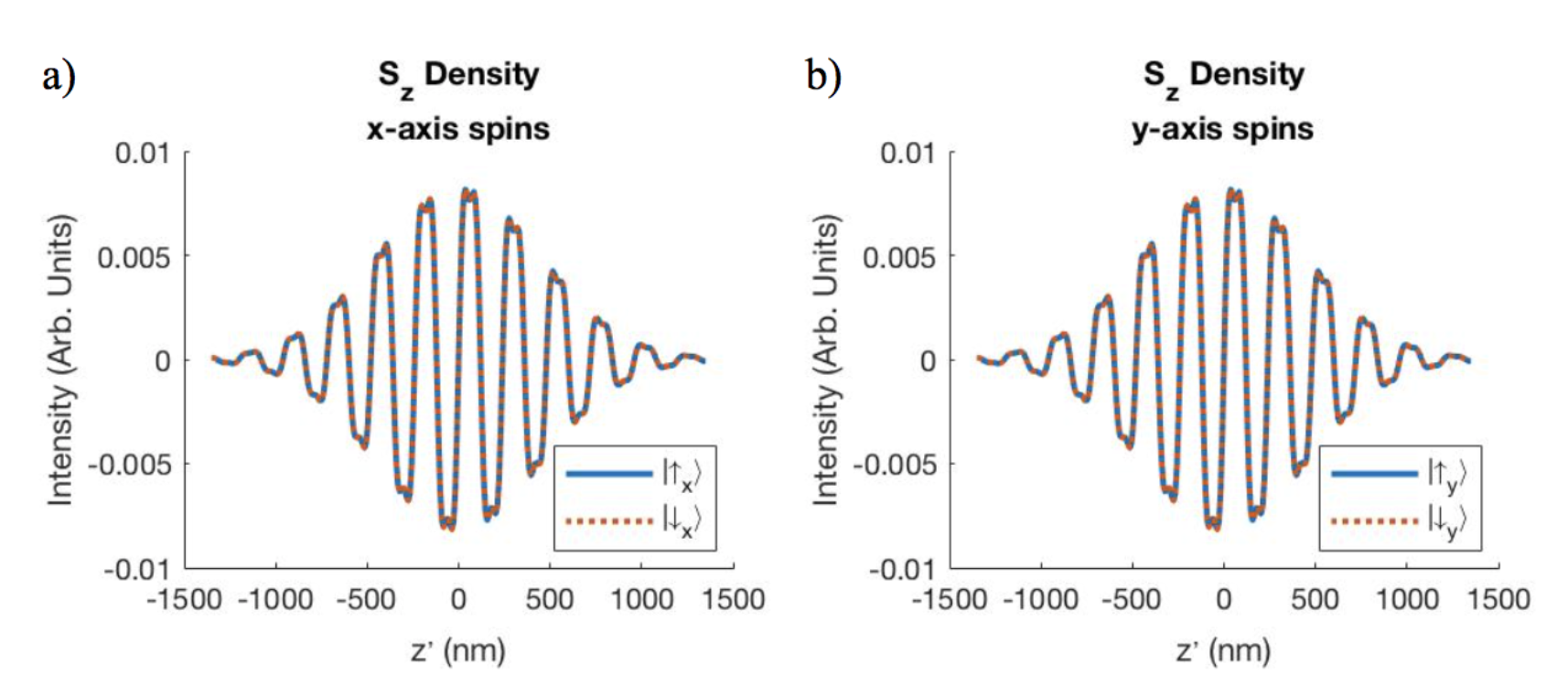}
    \caption{Shows the separated spin longitudinal spin densities for initially unpolarized beam quantized along the a.) x-axis, and b.) y-axis. Interaction with the longitudinally varying magnetic field, followed by a drift length redistributes longitudinal spin-density into specific phase bins, in both cases the up and down spin have equal contribution in the initial spin-state so the resulting spin density is the same.}
    \label{fig:setup}
\end{figure}

Due to the redistribution of longitudinal spin density into phase bins, interaction with the second stage phase-matched stage enables transformation into final states with the same $\left[\hat{S}_y\right]$ as expected. Figure S3 depicts the $S_y$ density for the given initial spins states, where the corresponding ensemble average of the initial mixed unpolarized state $\left[\hat{S}_y\right] \approx 0.7\hbar/2$.

\begin{figure}[h]
    \centering
    \includegraphics[width=0.49\textwidth]{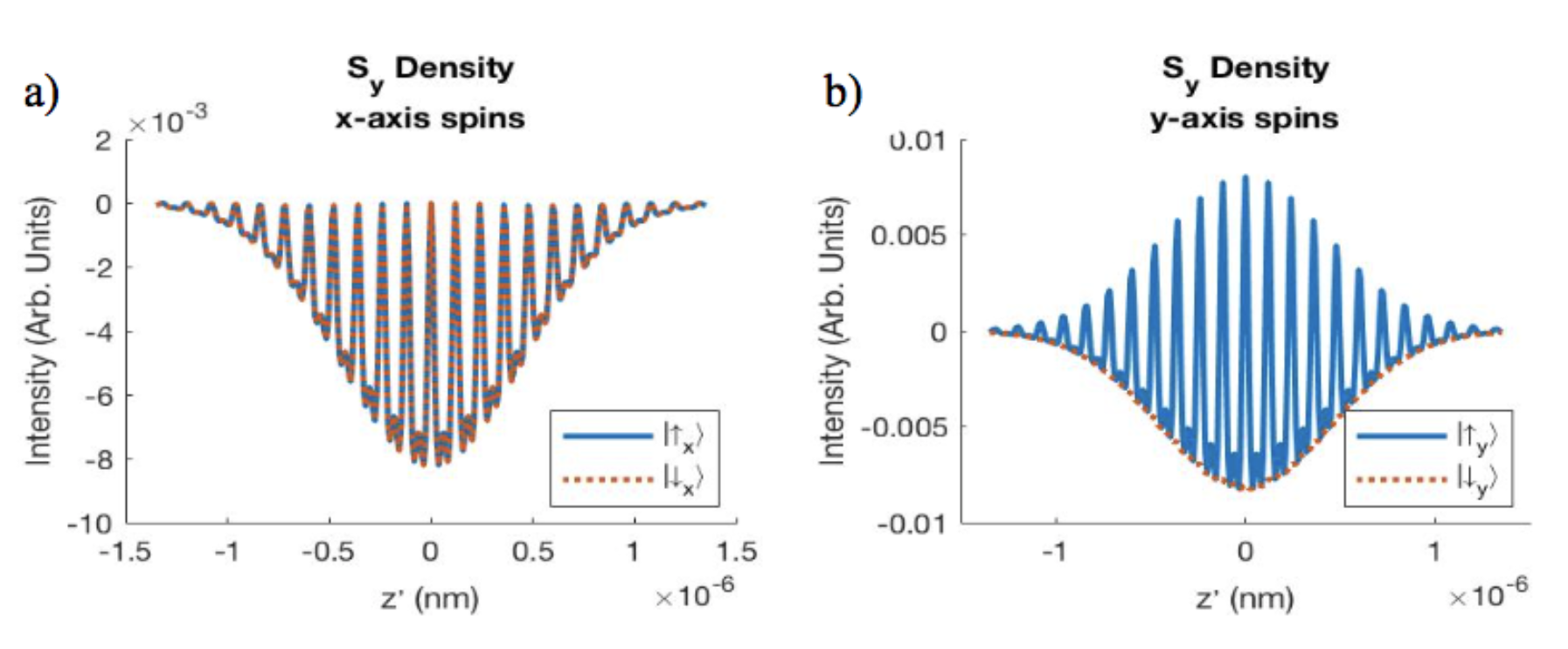}
    \caption{Depicts y-axis spin density after second-stage interaction from a.) up and down x-axis spins. The different superposition phases lead to overlap in final spin-state density. b.) Shows initial up and down y-axis spin states spin density, highlighting maximal asymmetry due to superposition phases. In this case one initial state results in very high, unity, polarization while the other results in lower polarization, with the ensemble average of the two states resulting in $\approx$70\% polarization.}
    \label{fig:setup}
\end{figure}

In general considering any arbitrarily large number of initial states that make up an unpolarized beam result in the same  ensemble average of y-axis spin expectation value. The first stage imparts a specific phase correlation to longitudinal spins, and the second stage imparts phase-dependent transformations that enables arbitrary unpolarized mixed states to be observed with the same degree of polarization. The specific wave-packet structure is dependent on initial spin states.

\bibliographystyle{apsrev}
\bibliography{references}